\documentclass[twocolumn]{aastex631}

\usepackage{subfigure}
\usepackage{graphicx}
\usepackage{times}
\usepackage{amsmath, amssymb}
\usepackage{url}

\usepackage[mathlines]{lineno}

\usepackage[]{xcolor}

\definecolor{myg}{cmyk}{0.75002,0,1,0}

\definecolor{msnote}{hsb:rgb}{0.492,0.492,0.492}

\begin{document} 

\title{Radio-X-ray Time Lags in GX 339-4: Probing Magnetic Field Transport in Black Hole Accretion}
%\title{Time lag between the radio and X-ray emission: Observational signatures of magnetic field transport around a black hole accretion disk}

\correspondingauthor{Bei You, Xinwu Cao}
\email{youbei@whu.edu.cn, xwcao@zju.edu.cn}

\author[0009-0000-9220-8543]{Dizhan Du}
\affiliation{Institute for Astronomy, School of Physics, Zhejiang University, Hangzhou 310058, China}
%\email{dudizhan@zju.edu.cn}

\author[0000-0002-8231-063X]{Bei You}
\affiliation{Department of Astronomy, School of Physics and Technology, Wuhan University, Wuhan 430072, China}

\author[0000-0002-5385-9586]{Zhen Yan}
\affiliation{Shanghai Astronomical Observatory, Chinese Academy of Sciences (CAS), Shanghai 200030, China}

\author{Yuao Ma}
\affiliation{Institute for Astronomy, School of Physics, Zhejiang University, Hangzhou 310058, China}

\author{Xinwu Cao}
\affiliation{Institute for Astronomy, School of Physics, Zhejiang University, Hangzhou 310058, China}

\begin{abstract}
We present an analysis of the time delay between the radio emission and the X-ray Compton luminosity during the 2010–2011 outburst of GX 339–4. Using the interpolated cross-correlation function (ICCF), we measure the time delay between the Compton luminosity and the radio luminosity, and find that during the rising hard state, the radio emission precedes the Compton luminosity by approximately 3 days. In contrast, in the decaying hard state, the radio emission lags behind the Compton luminosity by about 8 days. By estimating the mass accretion rate and the disk truncation radius, the calculated inner magnetic field can account for both the radio delay in the decaying hard state and the radio precedence in the rising hard state. The time delays observed in different outbursts across multiple sources are compared further, and the underlying physical mechanisms account for this difference are discussed. These results provide insights into the evolving coupling between the inner accretion flow and the jet in black hole X-ray binaries.
\end{abstract}

%\title{Multi-wavelength time-lags of MAXI J1348-630} 

\section{Introduction} \label{sec:intro}

A black hole X-ray binary (BHXRB) is a binary system consisting of a black hole (BH) and a companion star. As matter is accreted from the companion star, an accretion disk forms around the black hole. BHXRBs are categorized into persistent and transient systems. Persistent BHXRBs maintain continuously high X-ray fluxes with moderate variability, whereas transient BHXRBs undergo dramatic outbursts in X-ray, optical, and radio wavelengths. These outbursts of transient sources are often with long periods of quiescence, which can last for years, decades, or even longer \citep{mcclintock_black_2006}. %Emissions at different wavelengths likely originate from different emitting regions and radiative processes.

Soft X-ray radiation arises from the accretion flow near the black hole, where gravitational energy is dissipated through viscous processes \citep{ss1973}, while hard X-ray emission is primarily generated by the scattering of soft X-rays in the Comptonizing corona \citep{narayan_advection-dominated_1998, poutanen_accretion_1998, Kawamura2023, you2021, you_observations_2023, you_delayed_2024}, and/or from relativistic jets \citep{markoff_jet_2001, kara2019, ma2021, marino2021, peng2023, zhang2023}. In the optical bands, X-ray reprocessing, viscous heating, and jets have been suggested to explain the observed emission \citep{cunningham_returning_1976, van_paradijs_absolute_1994, van_paradijs_accretion_1996, russell_global_2006, corbel_near-infrared_2002, markoff_exploring_2003, brocksopp_soft_2004, john_correlated_2024, mastroserio_x-ray_2025, du_comprehensive_2025,fan2025}. Radio emission primarily originates from synchrotron radiation in relativistic jets \citep{fender_jets_2006, bright2020}. 

%High-resolution radio observations have revealed a relativistic jet launched near the black hole \citep{fender2004araa}. The Blandford–Znajek (BZ) and Blandford–Payne (BP) mechanisms have been proposed to power the jet \citep{blandford1977, blandford1982}, driven by the interplay between magnetic fields and the rotation of the black hole or accretion disk, respectively. Jets powered by either the BZ or BP mechanisms ultimately require a large-scale magnetic flux near the black hole. The buildup of magnetic fields through accretion remains a topic in black hole astrophysics \citep{tchekhovskoy2011}. Two main processes have been proposed for generating such coherent fields in the outer thin disk. The first is the inverse-cascade dynamo, in which magnetic loops in the outer disk connect to the inner ADAF and are advected inward toward the black hole \citep{tout_can_1996, beckwith_transport_2009}. The second, suggested by numerical simulations \citep{liska_large-scale_2020}, is turbulence in a thick disk that amplifies large-scale poloidal flux and can lead to the formation of a magnetically arrested disk (MAD), although this process is less efficient in thin disks. 

High-resolution radio observations have revealed a relativistic jet launched near the black hole \citep{fender2004araa}. The Blandford–Znajek (BZ) and Blandford–Payne (BP) mechanisms were proposed as the primary engines powering relativistic jets \citep{blandford1977, blandford1982}, driven by the interaction between magnetic fields and the rotation of the black hole, and the interaction between magnetic fields and the accretion disk, respectively. Either the BZ or the BP mechanism requires a large-scale magnetic flux near the black hole. The process by which magnetic fields build up through accretion remains an ongoing topic in black hole astrophysics \citep{tchekhovskoy2011}. Two main processes have been proposed for generating such coherent fields in the outer thin disk. The first is the inverse-cascade dynamo, in which magnetic loops in the outer disk connect to the inner ADAF and are advected inward toward the black hole \citep{tout_can_1996, beckwith_transport_2009,livio1999,lijw2019,cao2021}. The second, suggested by numerical simulations \citep{liska_large-scale_2020,davis_magnetohydrodynamics_2020,jacquemin-ide_magnetically_2019}, involves turbulence in a thick disk that amplifies large-scale poloidal flux and can lead to the formation of a magnetically arrested disk (MAD)\citep{narayan2003,igumenshchev2003,zamaninasab2014,yuan2022}, although this process is less efficient in thin disks.

To understand the transport of the magnetic field onto black holes, multi-wavelength observations from radio to X-rays and $\gamma$-rays are essential \citep{you_observations_2023}. Broadband spectral energy distribution (SED) fitting \citep{marino2021,rodi2021,ozbey2022,banerjee2024,et2024,yoshitake2022} and radio/X-ray correlations \citep{corbel2003,coriat2011,williams2022} have been widely used to probe the origin of the emission and the underlying accretion–ejection physics. In correlation studies, radio and X-ray luminosities are often compared, showing a non-linear relation $L_{\rm radio} \propto L_{\rm X}^{0.6-0.7}$ during the low hard state \citep[][]{gallo2003,corbel_radiox-ray_2003,corbel_formation_2013,carotenuto2021/505}. \cite{jiang_physical_2024} proposed a model where the magnetic field from the outer disk is advected inward and enhanced near the black hole, launching relativistic jets via the Blandford–Znajek mechanism. This leads to an increase in both jet power and X-ray emission with the mass accretion rate, matching the observed radio/X-ray correlation. Spectral decomposition further reveals the evolution of individual emission components \citep{you2021,fijma2022,dai2023,peng2023}. More recently, multi-wavelength spectral–timing studies have provided insights into fast variability from jets in BHXRBs \citep{Tetarenko2019,Tetarenko2021b,yang2025}, identifying X-ray–radio correlations and short timescale time-lags between optical, X-ray, and radio bands, which place key constraints on jet composition.

\cite{you_observations_2023} investigated the multi-wavelength behavior of MAXI J1820+070 during its 2018 outburst and discovered that the radio emission lagged the X-ray Compton emission by $\sim$8 days in the decaying hard state—the longest delay ever observed in low-mass BHXRBs. This delay traces the launching and suppression of the jet, driven by magnetic fields dragged from the disk into the inner advection-dominated accretion flow (ADAF) near the BH \citep{cao2011,yuan2014,2023ApJ...944..182D}. The unprecedented lag was interpreted as a consequence of the ADAF expansion (i.e., recession of the thin disk in the truncated-disk scenario), during which the magnetic field is advected inward and continuously amplified, even after the hard X-ray peak. 
However, in the rising hard state, no time delay was observed in the radio and X-ray Compton emission. 

During the 2019 outburst of MAXI J1348-630, \cite{you_delayed_2024} reported a shorter lag of about 3 days between the radio and X-ray Compton fluxes, specifically during the rising hard state. Notably, during the decaying phase and the subsequent mini-outburst, the jet reappears, and the Compton and radio emissions become nearly simultaneous.
%At the radio maximum, about 8 days after the X-ray peak, the accumulated field becomes dynamically important, with magnetic forces comparable to gravity at the inner edge \citep{narayan2003}. Thus, \cite{you2023} provided the first observational evidence for the formation of a magnetically arrested disk (MAD) in a BH accretion system.

%Although both are classified as BHXRBs, they display markedly different radio–X-ray Compton time-lag behaviors. To understand the underlying accretion and magnetic mechanisms responsible for these difference, it is crucial to expand the sample size and assess whether a universal trend exists.
Although both MAXI J1820+070 and MAXI J1348-630 are classified as low-mass BHXRBs, they exhibit markedly different radio/X-ray time-lag behaviors, with variations also seen in other sources (see below), such as H1743-322 \citep{mcclintock_2003_2009} and GRS 1739-278 \citep{hjellming_imaging_1997}. To understand the underlying accretion and magnetic mechanisms responsible for these differences—potentially tied to the evolution of truncation radii and mass accretion rates—it is crucial to expand the sample size and assess whether a universal trend exists across BHXRBs.

GX~339$-$4, which hosts a black hole with a mass of \(2.3-9.5\,M_{\odot}\) in a binary system with a low-mass companion, located at a distance of \(>5\,\mathrm{kpc}\) \citep{hynes_distance_2004, heida_mass_2017}, is a well-known BHXRB extensively monitored in radio and X-ray during its 2010-2011 outbursts, including both the rising and decaying hard states \citep{miyamoto_x-ray_1991, corbel_coupling_2000, corbel2003, corbel_universal_2013}. 
These high-cadence observations make GX 339-4 an ideal source for studying the time evolution of the accretion flow and magnetic field near the black hole, 
particularly through time lags that probe advection and amplification processes. 

In this study, we investigate the time delays between radio emissions and X-ray Compton during the rising and decaying hard states of GX~339$-$4. Section~\ref{sec:data} describes our procedures for data reduction, model selection, spectral fitting, and calculations of X-ray luminosity. In Section~\ref{sec:time ccf}, we present the ICCF method employed to quantitatively analyze the time delays between radio and X-ray emissions, along with our findings. Finally, the implications of our results are discussed and physically interpreted in Section~\ref{sec:dis}. Our findings are summarized in Section~\ref{sec:sum}.

%\section{data reduction and analysis} \label{sec:data}
\section{Data reduction and X-ray spectral analysis} \label{sec:data}

In this study, we aim to investigate the time delays between radio emission and X-ray Compton. Due to the limited radio observation cadence (only the 2010-2011 outburst had sufficient coverage), we focus solely on that outburst.
In this outburst, we analyze \textit{Rossi X-ray Timing Explorer (RXTE)} archival observations of GX 339-4 collected between MJD 55200 and MJD 55700. 
Good time intervals for each observation are generated using the \texttt{maketime} task. The source and background spectra are extracted with the \texttt{pcaextspec2} script, and the spectra are grouped to ensure a minimum of 25 counts per bin.

\begin{figure*}
\centering
\includegraphics[width=\textwidth]{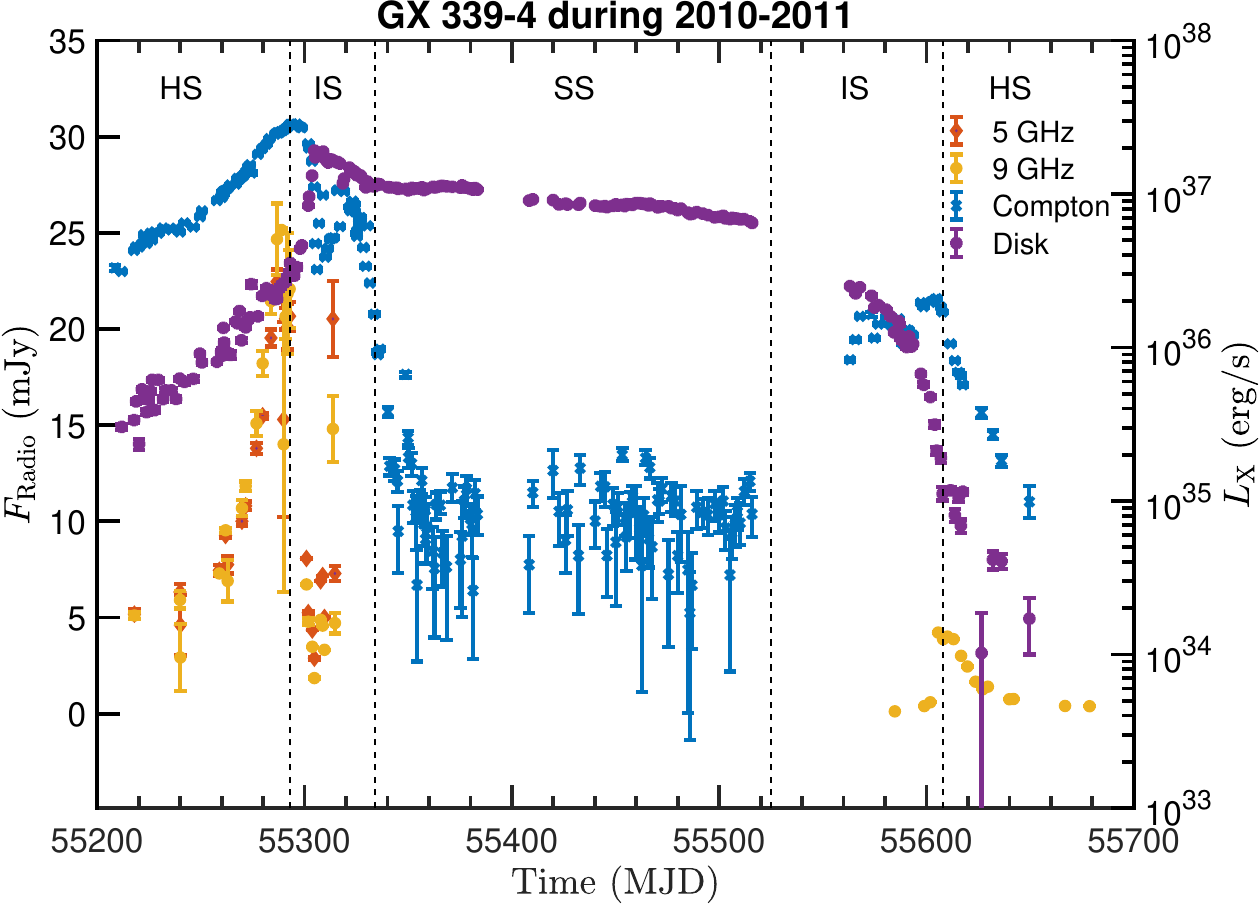}
\caption{
The radio and X-ray lightcurves of GX 339-4 from MJD 55200 to 55700 , with the dashed lines indicating the transitions between the hard state (HS), intermediate state (IS), and soft state (SS).
}
\label{fig:lightcurve}
\end{figure*}

To analyze the energy spectra during the GX 339-4 outbursts in the 3–25 keV range, we used the software package \texttt{XSPEC} (version 12.10.1), applying a 1\% systematic error to the data. The details of the spectral fitting are provided in \cite{du_comprehensive_2025}.
The light curves are shown in Fig.~\ref{fig:lightcurve}. 
A significant rise in the disk component is observed at the beginning of the outburst. It peaks around MJD 55300, increasing by roughly two orders of magnitude, and then gradually declines until MJD 55520. Subsequently, it enters a rapid decay phase, during which the disk luminosity decreases by about two orders of magnitude within $\sim100$ days.

The rise in the Compton component closely tracks that of the disk component until MJD 55300, when the Compton emission peaks, after which it declines sharply, with a minor flare near MJD 55320. It subsequently remains at a low luminosity level—approximately two orders of magnitude below its peak, until MJD 55520. After this epoch, the Compton component undergoes a reflare, peaking at MJD 55600, followed by a rapid decline. By around MJD 55700, the outburst returns to quiescence. 

The X-ray spectra of BHXRBs exhibit significant evolution during outbursts, typically displaying two primary shapes: dominance by a hard power-law component in the hard state or by a soft blackbody-like component peaking at a few keV in the soft state. Sources like GX 339-4 begin in the hard state during the rising phase and transition to the soft state, while the reverse transition (soft to hard) occurs at lower luminosities, reflecting hysteresis in the hardness-intensity diagram \citep{remillard_x-ray_2006}. Intermediate states during these transitions are recognized as distinct phases with unique spectral and timing properties \citep{homan_evolution_2005, nandi_accretion_2012}. For this work, we adopt the spectral state classifications from \cite{marcel_unified_2020}, based on RXTE/PCA continuum fits from \cite{clavel2016} for the 2010-2011 outburst, without distinguishing between hard-intermediate and soft-intermediate states: hard state (MJD 55208–55293 rising, MJD 55608–55646 decaying), soft state (MJD 55334–55525), and intermediate states (MJD 55293–55334, MJD 55525–55608).

In the radio band, GX 339$-$4 was monitored by the Australia Telescope Compact Array (ATCA) at 5.5 GHz and 9 GHz \citep{corbel_universal_2013,corbel_formation_2013}. The observations at 9 GHz span both the rising and decaying phases of the hard and intermediate states, whereas the 5.5 GHz data cover only the rising phases of the hard and intermediate states.
%The evolution of the spectral index indicates that $\alpha \sim 0$ after MJD 58629. So we estimate the spectral index $\alpha$ of MeerKAT data as 0 during the mini-outburst. 
The evolution of the estimated radio fluxes is plotted in Fig. \ref{fig:lightcurve}.
The radio lightcurves broadly resembles that of the X-ray Compton component. Radio emission is detected at the onset of the outburst at a level of $\sim5$ mJy. As the system brightens in the hard state, the radio emission rises rapidly, reaching $\sim25$ mJy at MJD 55288. During this phase, the increasing radio emission is consistent with the presence of compact jets.

In the intermediate state, a sharp drop in the radio flux density is observed between MJD 55288 to 55310, during which the 9-GHz flux density decreases from $\sim25$ to $\sim2$ mJy. The radio emission then shortly rebrightens to $\sim15$ mJy at MJD 55313, coincident with enhanced X-ray Compton emission. This behavior may be related to episodic jet activity \citep{yuan_magnetohydrodynamical_2009}, which is beyond the scope of this work.
In the soft state, where compact jet radio emission is generally not expected for BHXRBs. After MJD 55580, the radio emission brightens again and subsequently decays.

\begin{figure*}[ht]
	\centering
    \includegraphics[width=0.95\textwidth]{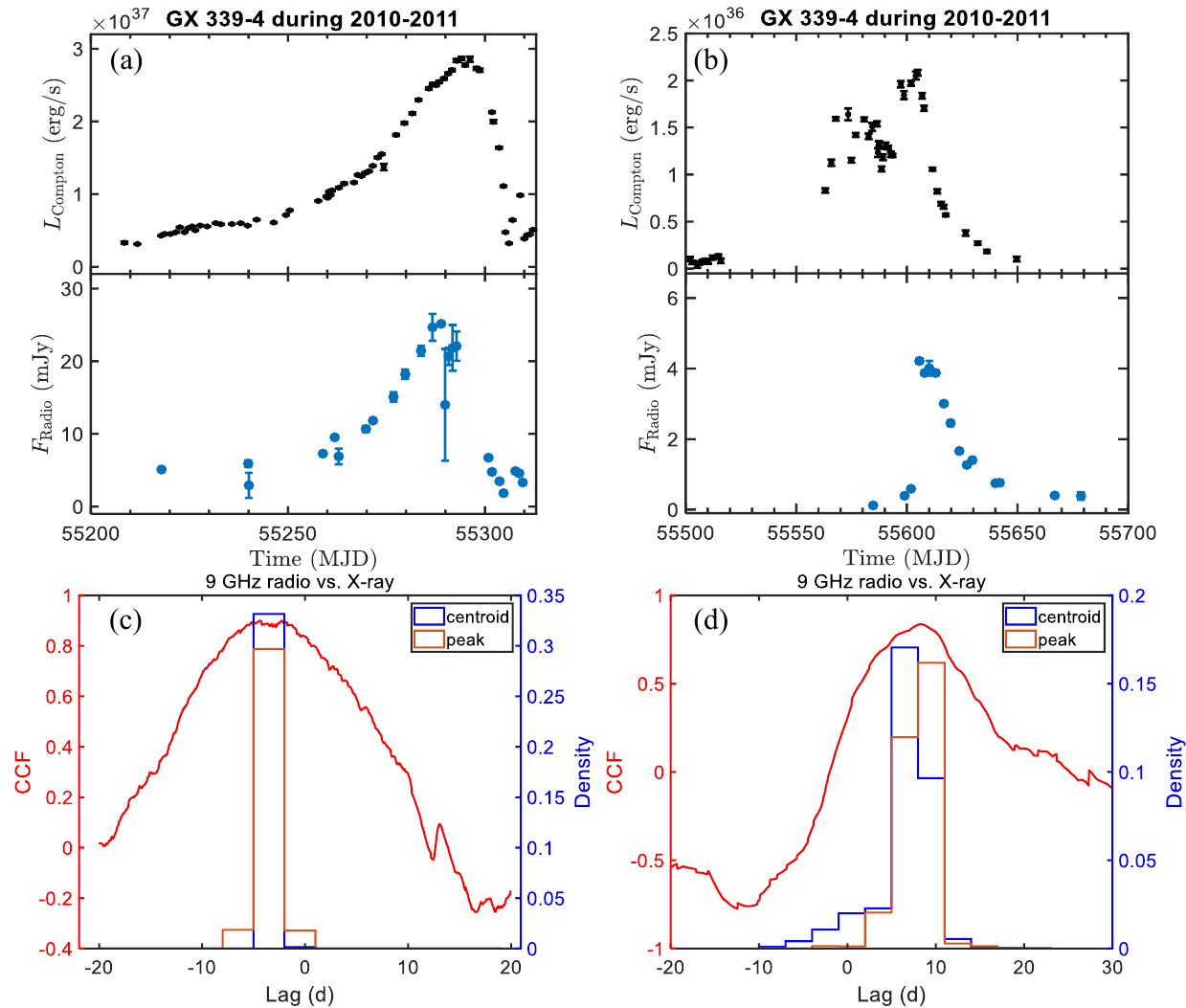}
	\caption{Panel a: Radio and Compton X-ray monitoring during the rising hard state flare. Panel b: Radio and X-ray Compton emission monitoring during the decaying hard state flare. Panel c: The cross-correlation analysis between the 9 GHz radio and Compton X-ray luminosity, specifically before MJD 55350 (red line). Panel d: Cross-correlation analysis between 9 GHz radio and the Compton X-ray luminosity, after MJD = 55500 (red line). In the lower panels, the blue histograms show the distribution of cross-correlation centroid lags, determined from the centroid of ICCF above a threshold (\(r>0.8r_{\rm max}\), where $r_{\rm max}$ is the maximum correlation coefficient). The brown histograms display the peak lags, corresponding to the time delays at which the maximum correlation occurs. The uncertainties in the lags are derived using the flux randomization/random subset sampling (FR/RSS) method. The corresponding axes are shown on the right. We performed 10,000 simulations using linear interpolation; details can be found in \citet{gaskell_accuracy_1987} and \citet{du_supermassive_2014}.}
	\label{fig:339ccf}
\end{figure*}

\section{Time-delay between the X-ray and radio emission}\label{sec:time ccf}

The interpolated cross-correlation function (ICCF) is a powerful tool for analyzing time delays between time-series signals, widely used in reverberation mapping (RM) of active galactic nuclei (AGNs) \citep{kaspi_reverberation_2000, bentz_lick_2009}. In this work, we use ICCF to investigate the coupling between accretion flow and jets by measuring time lags between radio and hard X-ray fluxes. Details of the ICCF implementation in BHXRBs are provided in \cite{you_observations_2023}.

In both the rising and decaying hard states, the Compton flux and the radio flux exhibit pronounced flares, whereas the disk component shows only a monotonic increase or decrease. 
During the rising hard state, the Compton component consistently lags the 9~GHz radio emission by several days. The cross-correlation analysis between the 9 GHz radio and Comptonized X-ray luminosities was carried out using data prior to MJD 55350. The centroid delay is measured via ICCF to be $\Delta \tau_{\rm rx}^{\rm c} = 3.58^{+0.55}_{-0.46}$~days, and the peak delay is $\Delta \tau_{\rm rx}^{\rm p} = 3.68^{+1.40}_{-1.16}$~days (Fig.~\ref{fig:339ccf}c). The results at 5.5~GHz are similar.
We further carry on the ICCF analysis on the radio/X-ray lightcurves during the decaying hard state (after MJD 55500). It turns out that the 9~GHz radio emission lags behind the Compton component by several days, The measured centroid delay is $\Delta \tau_{\rm rx}^{\rm c} = 7.43^{+1.51}_{-1.52}$ days, and the peak delay is $\Delta \tau_{\rm rx}^{\rm p} = 8.18^{+0.66}_{-2.40}$ days (Fig.~\ref{fig:339ccf}d).

\section{Discussion}\label{sec:dis}

In previous sections, we studied the time delays between the jet radio emission and the X-ray Compton emission during the 2010-2011 outburst of GX 339-4.
In the following, we use the time delays between the jet radio emission and the X-ray Compton emission to study the time evolution of the accretion flow and the magnetic field during the outburst of this source. 
%The quantitative modeling and the interpretation of these observed radio-X-ray delays will be comprehensively presented in our next work.     
\subsection{Delayed radio emission during the decaying hard state}\label{sec:d_discuss}

As shown in Fig.~\ref{fig:339ccf}b, during the decaying state, both the flares of the radio emission and X-ray Compton emission were observed. The X-ray Compton emission peaks around MJD 55604, after which it fades toward the quiescent state. Surprisingly, after the peak of X-ray Compton emission, the radio emission continues to increase, peaking around MJD 55610. The ICCF analysis reveals that the X-ray Compton emission lags the radio emission by approximately 7 days. This delay in radio emission is significantly longer than what would be expected from jet travel alone\citep{fender_powerful_2001,fender_towards_2004}.

During the 2018 outburst of MAXI~J1820$+$070, the radio fluxes were observed to be delayed by approximately 8 days in comparison to the hard X-ray emission \citep{you_observations_2023}. This finding aligns with the results of our study. In their interpretation, as the accretion rate decreases, the truncation radius \(R_{\rm tr}\) expands. This expansion allows the ADAF to grow and intensify the magnetic field near the black hole. The increased strength of the magnetic field enhances the jet power, resulting in radio emission peaking approximately 8 days after the hard X-ray emission peak.

The observed radio emission originates from the jets, depending on the magnetic field of the accretion. 
%Jets are powered by the electromagnetic extraction of rotational energy from the black hole, a process known as the BZ mechanism \citep{blandford1977}.
To estimate the evolution of the jet power, we computed the magnetic field using the method employed by \citet{you_observations_2023} and \citet{cao2011}.

The large-scale magnetic field originating from the outer thin disk is advected and diffused inward, leading to a substantial enhancement of the field strength in the ADAF \citep{cao2011}. The magnetic field $B_{\rm tr}$ near the truncation radius $R_{\rm tr}$ of the outer thin disk is estimated as \citep[equation~27 in][]{cao2021}:
\begin{equation}\label{equ:b_pd_gas}
B_{\rm tr}\sim 2.48\times 10^{8} \alpha^{-1/20}m^{-11/20}\dot{m}_{\rm d}^{3/5} R_{\rm tr}^{-49/40}~{\rm Gauss}, 
\end{equation}
where we assume $m = M_{\rm BH}/M_{\odot} = 6$ \citep{hynes_distance_2004}. Since our focus is on the time delay, the uncertainty in $m$ affects only the absolute values and does not influence the delay itself. The parameter $\alpha$ denotes the viscosity parameter, for which we adopt a typical value of 0.2 \citep{osaki_accretion_1974}.
The truncation radius $R_{\rm tr}$ and the Eddington-scaled mass accretion rate, $\dot{m}_{\rm d} = \dot{M}_{\rm d}/\dot{M}_{\rm Edd}$, as functions of time are taken from the recent estimates for the 2010--2011 outburst of GX~339$-$4 reported by Xu et al. (in preparation), and are used in our subsequent calculations.

To estimate the enhancement of the magnetic field strength in the ADAF at different times, we adopt the method described in Section~2.2 of \cite{cao2011}, which involves several key parameters. The magnetic Prandtl number, $P_{\rm m} \equiv \nu / \eta_{\rm m}$, where $\nu$ is the turbulent viscosity and $\eta_{\rm m}$ is the magnetic diffusivity, is found to be around unity, or $\sim 2 - 5$ from shearing box simulations \citep{parker1979, fromang2009, guan2009, lesur2009}, with a value of 0.75 adopted in our study. The parameter $H/R$ (the height-to-radius ratio) was assumed to be 0.8 in the calculations \citep{narayan2022, begelman2022}, and the innermost stable circular orbit was set to $R_{\rm ISCO} = 4R_{\rm g}$ \citep{cao2011}, where $R_{\rm g} = GM_{\rm BH}/c^2$ is the gravitational radius. The relative importance of the outflows on the radial velocity was described by the value of $f_m$, where the term $\sim 1/(1 + f_{\rm m})$ represents the fraction of kinetic energy in the accretion flow that is radiated away, while the remaining fraction, $\sim f_{\rm m}/(1 + f_{\rm m})$, is used to accelerate the outflows \citep{cao_accretion_2016}.
In this work, we set $f_m = 1.15$, obtained by matching equation~(\ref{equ:lcom}) to the observed values\citep{you_observations_2023}.

%The value of $p_{\rm w} = 0.01$ is adopted to describe the radial dependence of the ADAF accretion rate due to outflows \citep{xie_radiative_2019}.

\begin{figure}[ht]
	\centering
    \includegraphics[width=1\linewidth]{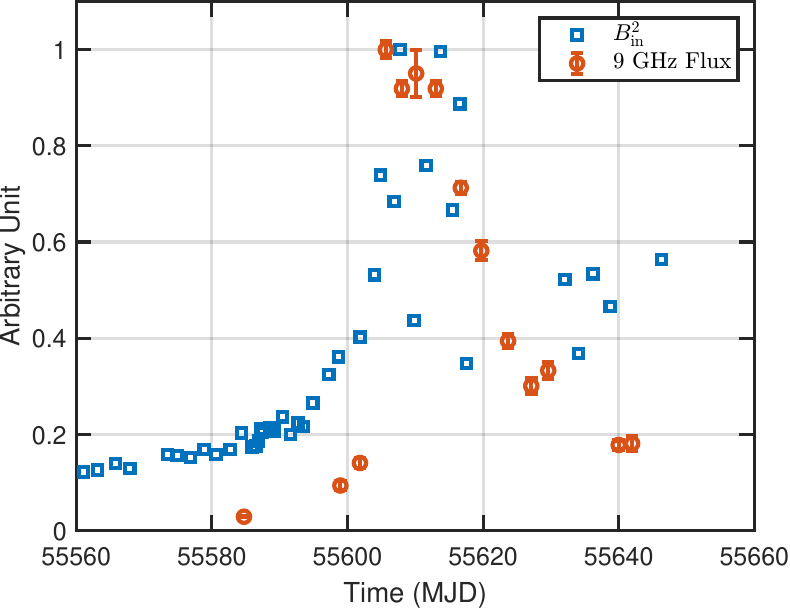}
	\caption{Squared inner magnetic field, $B_{\rm in}^2$, and 9 GHz radio flux in arbitrary units during the decaying hard state flare.}
	\label{fig:339_d_br}
\end{figure}

By calculating the magnetic field at different times, and noting that the jet power is proportional to the square of the inner magnetic field (magnetic field at innermost stable circular orbit), $B_{\rm in}^2$, we present the temporal evolution of $B_{\rm in}^2$ in Fig.~\ref{fig:339_d_br}. The evolution closely matches the observed radio time delays.

 \subsection{Preceding radio emission during the rising hard state}\label{sec:r_discuss}

As shown in Fig.~\ref{fig:339ccf}a, during the rising phase, the onsets of both the radio emission and X-ray Compton emission were observed around MJD 55200. 
Then, the radio and X-ray Compton emissions rise. The radio emission first peaks around MJD 55286 and then fades toward the hard-to-soft state transition. Surprisingly, after the radio peak, the X-ray Compton emission continues to increase with time and appears to peak around MJD 55294. The ICCF analysis reveals that the X-ray Compton emission lags the radio emission by about 3 days.
To explain the early radio emission observed during the rising hard state, we also attribute this behavior to variations in the strength of the inner magnetic field over time. A similar approach to that used for the decaying hard state was applied: we first estimate the time-evolution of the mass accretion rate.

As for the rising hard state, 
%is considered the inverse of the decay phase. In this framework, 
%we assume that the mass accretion rate follows an exponential rise toward a peak, and then is followed by an exponential decay.
%Unlike the decaying hard state, for which the declining mass accretion rate can be extrapolated from the disk luminosity observed in the soft state, as done by Xu (2025, in preparation), 
the following equation for estimating the Compton luminosity is used to constrain the Eddington-scaled accretion rate, $\dot{m}$.
\begin{equation}
\lambda_C(t) \equiv \frac{L_C(t)}{L_{\rm Edd}} \sim \frac{\dot{m}(t) - \lambda_d(t)}{1 + f_m} \left[\frac{2.25 R_{\rm ISCO}}{R_{\rm tr}(t)}\right]^{p_{\rm w}},
\label{equ:lcom}
\end{equation}
\citep[see equation~S5 in][]{you_observations_2023}.
Here, $L_C$ is the Eddington-scaled Compton luminosity, and $\lambda_d \equiv L_d / L_{\rm Edd}$ denotes the Eddington ratio of the observed disk luminosity. The quantity $\dot{m}$ is the Eddington-scaled mass accretion rate. The condition $\dot{m}(t) = \lambda_d$ applies only when the disk is not truncated, i.e., when $R_{\rm tr} = R_{\rm ISCO}$. \( R_{\rm ISCO} \) is the radius of the innermost stable circular orbit (ISCO). We adopt $p_{\rm w} = 0.01$ to describe the radial dependence of the ADAF accretion rate due to outflows \citep{xie_radiative_2019}, and use $f_m = 1.15$, the same value as in the decaying hard state (see Sect.~\ref{sec:d_discuss}).

The truncation radius $R_{\rm tr}$ can be estimated using the following equation
\begin{equation}\label{equ:rtr}
\frac{R_{\rm tr}(t)}{R_{\rm ISCO}} \simeq \frac{\dot{m}(t) \lambda_{\rm d}(t_0)}{\dot{m}(t_0) \lambda_{\rm d}(t)}
\end{equation}
\citep[see equation~S2 in][]{you_observations_2023}. \( t_0 \) correspond a arbitrary epoch when the source is in the soft state, at which the truncation radius \( R_{\rm tr} \) reaches \( R_{\rm ISCO} \), and \( \lambda_{\rm d}(t_0)/\dot{m}(t_0) = 1 \). 

The observed disk and Compton luminosities are substituted into Eq.~\ref{equ:lcom} to derive the mass accretion rate. To obtain an analytical expression for subsequent calculations, we fit the inferred mass accretion rate with a broken exponential, i.e., a function consisting of an exponential rise and decay. The resulting fit is shown as follows,
\begin{equation}
\dot{m}(t) =
\begin{cases}
7.3 \times 10^{-3} \exp\left[\dfrac{t - 55200}{40.5}\right], & t < 55294, \\[0.6em]
9.98 \times 10^{-1} \exp\left[-\dfrac{t - 55200}{37.5}\right], & t \ge 55294.
\end{cases}
\label{equ:mdot}
\end{equation}

%We also adopt the method of calculating the intensity of the inner magnetic field, as described in Section~\ref{sec:d_discuss}, using the same parameters. The squared magnetic field, $B_{\rm in}^2$, which changes over time, is shown in Fig.~\ref{fig:339_r_br}(a), with data from three specific epochs highlighted. The corresponding magnetic field intensity variations with radius are depicted in Fig.~\ref{fig:339_r_br}(b). The results indicate that the calculated magnetic field effectively accounts for the observed radio precede.

\begin{figure}[ht]
	\centering
    \includegraphics[width=1\linewidth]{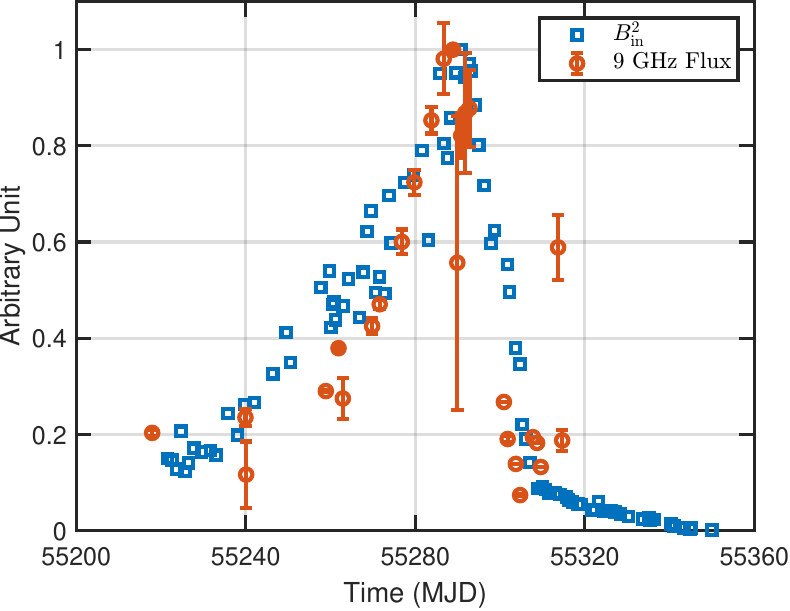}
	\caption{Squared inner magnetic field, $B_{\rm in}^2$, and 9 GHz radio flux in arbitrary units during the rising hard state flare.}
	\label{fig:339_r_br}
\end{figure}

The method outlined in Sect.~\ref{sec:d_discuss} is adopted to calculate the strength of the inner magnetic field, using the same set of parameters. The squared inner magnetic field, \(B_{\rm in}^2\), which varies over time, is illustrated in Fig.~\ref{fig:339_r_br}. The results indicate that the calculated magnetic field successfully matches the temporal evolution of the observed radio emission.

 \subsection{Comparison of the Radio/X-ray lag with previous studies}
 
In the previous sections, we reported that the Compton luminosity lags behind the radio emission during the rising hard state, whereas the radio emission lags behind the Compton luminosity during the decaying hard state. However, in other BHXRBs, the time delays do not always follow the same pattern as in GX~339$-$4. 

In MAXI J1820+070, a radio delay of approximately 8 days relative to the Comptonization was observed during the decaying hard state, but no time delay was observed during the rising hard state\citep{you_observations_2023}. 
In MAXI J1348-630, during the decaying phase and the subsequent mini-outburst, the Compton and radio emissions become nearly simultaneous. However, in the rising hard state, radio emission lagged the X-ray Compton component, which is the opposite of GX 339-4 \citep{you_delayed_2024}.

During the 2003 outburst of H1743-322, two flares occurred in the rising phase. In the first flare, the radio emission preceded the X-ray Compton emission, while in the second flare, the radio emission lagged behind the X-ray Compton emission \citep[see Figs. 13-14 in][]{mcclintock_2003_2009}. However, during the failed 2018 outburst, the radio and X-ray Compton appeared to occur simultaneously \citep[see Fig. 1 in][]{williams_2018_2020}.
The 2008 and 2009 outbursts of H1743-322 were observed with high cadence in both the radio and X-ray bands. However, the 2008 outburst lacked a clear rising part for both the radio and X-ray Compton, making it difficult to determine the time delay \citep{jonker_following_2010, coriat2011}. Similarly, the 2009 outburst also lacked a clearly defined rising phase, and the radio flare around MJD 54989 does not appear to be associated with a compact jet \citep{miller-jones_disc-jet_2012, coriat2011}.
In GRS 1739-278, the radio lagged behind the X-ray in the rising phase \citep[see Fig. 5 in][]{hjellming_imaging_1997}.

In order to understand the mechanism behind the
Given the diverse time-delay behaviours observed in different X-ray binaries, it is essential to carefully examine the dependence of both the Compton component and the radio emission (i.e., the magnetic field).
%Given that the radio intensity is directly related to the magnetic field, the scenario discussed in Sects.~\ref{sec:d_discuss} and \ref{sec:r_discuss} can also be used to explain the diverse time-delay behaviours as mentioned above.
%To investigate the physical origin of the distinct time-lag behaviors observed across different outbursts, 
The Compton luminosity and magnetic field are governed by two time-dependent parameters: the mass accretion rate $\dot{m}(t)$ and the truncation radius $R_{\rm tr}(t)$. 
By rewriting Eq.~(\ref{equ:rtr}) as
\begin{equation}
\lambda_d(t) = \frac{\dot{m}(t)}{R_{\rm tr}(t) / R_{\rm ISCO}},
\end{equation}
and substituting it into Eq.~(\ref{equ:lcom}), we obtain the Compton luminosity as
\begin{equation}
\lambda_C(t) \propto \dot{m}(t) \left[\Re(t)^{-p_{\rm w}} - \Re(t)^{-1-p_{\rm w}} \right],
\end{equation}
where $\Re ={R_{\rm tr}(t) / R_{\rm ISCO}}$.

\begin{figure}[ht]
	\centering
    \includegraphics[width=1\linewidth]{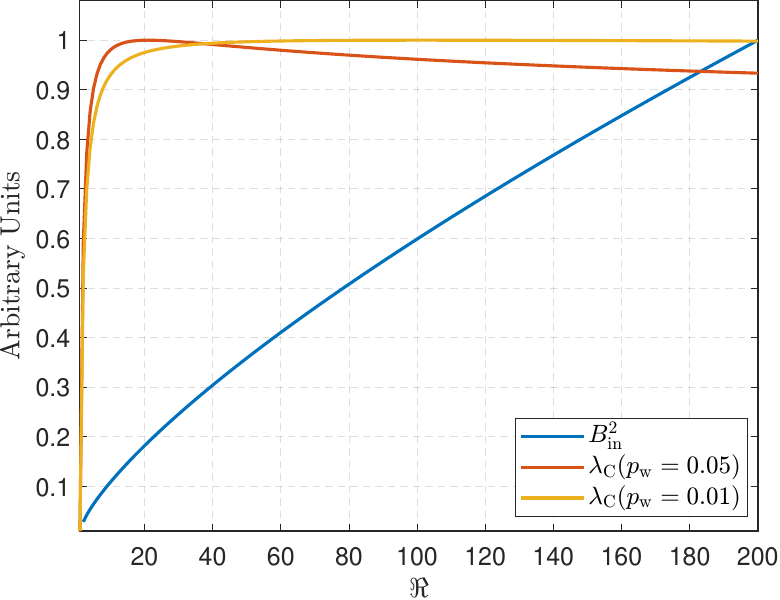}
	\caption{The variation of $B^2_{\rm in}$ and $\lambda_C$ as a function of $r_{\rm tr}$. The accretion rate, $\dot{m}_{\rm d}(t)$, is fixed at 0.1, and $p_{\rm w}$ is set to 0.05 and 0.01 as examples, with all other parameters adopted as described in Sect.~\ref{sec:d_discuss}.
}
	\label{fig:mag_f}
\end{figure}

The inner magnetic field $B_{\rm in}$ cannot be expressed in a simple analytical form directly. Although $B_{\rm in}$ is approximately proportional to $\dot{m}^{3/5}$ (see Eq.\ref{equ:b_pd_gas}), its dependence on $\Re$ can only be obtained through numerical simulations. We note that the magnetic field enhanced by the advection and diffusion processes within the ADAF is independent of the accretion rate $\dot{m}$ and is governed solely by the truncation radius $\Re$ \citep{cao2011}. Its dependence on $\Re$ is well characterized by a power law with an index of 0.37, which is derived by fitting the numerical calculations of \cite{cao2011}. Therefore, the inner magnetic field can be expressed as a function of $\Re$ and $\dot{m}$ as follows:
\begin{equation}
B^2_{\rm in} \propto \dot{m}(t)^{6/5}  \Re(t)^{0.74}.
\label{equ:bin2}
\end{equation}
The variation of $B^2_{\rm in}$ and $\lambda_C$ as a function of $\Re$ is presented in Fig.~\ref{fig:mag_f}, where $\dot{m}(t)$ is fixed at 0.1 and $p_{\rm w}$ is set to 0.05 and 0.01 to illustrate their dependence.

%We found that the Compton luminosity increases with the accretion rate $\dot{m}_{\rm d}(t)$, while the inner magnetic field follows a scaling relation of $B^2_{\rm in} \propto \dot{m}^{6/5}$. 
%More importantly, for a given accretion rate, the Compton luminosity exhibits a non-monotonic dependence on the truncation radius: it increases rapidly for $r_{\rm tr} < 1 + 1/p_{\rm w}$ and decreases gradually when $r_{\rm tr} > 1 + 1/p_{\rm w}$.
%Furthermore, the dependence of $B^2_{\rm in}$ on variations in $r_{\rm tr}$ is found to be significantly weaker than that of the Compton luminosity.

In the decaying hard state, as $\dot{m}$ decreases over time, the ADAF radius $\Re$ expands rapidly (Xu et al., in preparation; \citealp{you_observations_2023}). This rapid expansion could lead to a rebrightening of both the Compton luminosity and the magnetic field \citep{you_observations_2023}. 

However, the Compton luminosity exhibits a non-monotonic dependence on the truncation radius: it increases rapidly for $\Re < 1 + 1/p_{\rm w}$ and decreases gradually when $\Re > 1 + 1/p_{\rm w}$.
%Since $\lambda_{\rm C}$ rises rapidly with increasing $\Re$ for $\Re \lesssim 20$ (see Fig.\ref{fig:mag_f}), 
Once $\Re \gtrsim 20$, the decay of $\dot{m}$ dominates the evolution of the Compton luminosity, rendering the effect of further $\Re$ expansion negligible (see Fig.\ref{fig:mag_f}). The dominance of $\dot{m}$ in governing the evolution of $B_{\rm in}$ occurs later than that of the Compton luminosity. This leads to a delayed decline of the magnetic field relative to the Compton emission, ultimately producing the observed radio lag during the decaying hard state. The process described above can successfully account for the delay between the radio and X-ray emission in GX~339–4 and MAXI~J1820+070. 

\begin{figure*}
	\centering
    \includegraphics[width=1\linewidth]{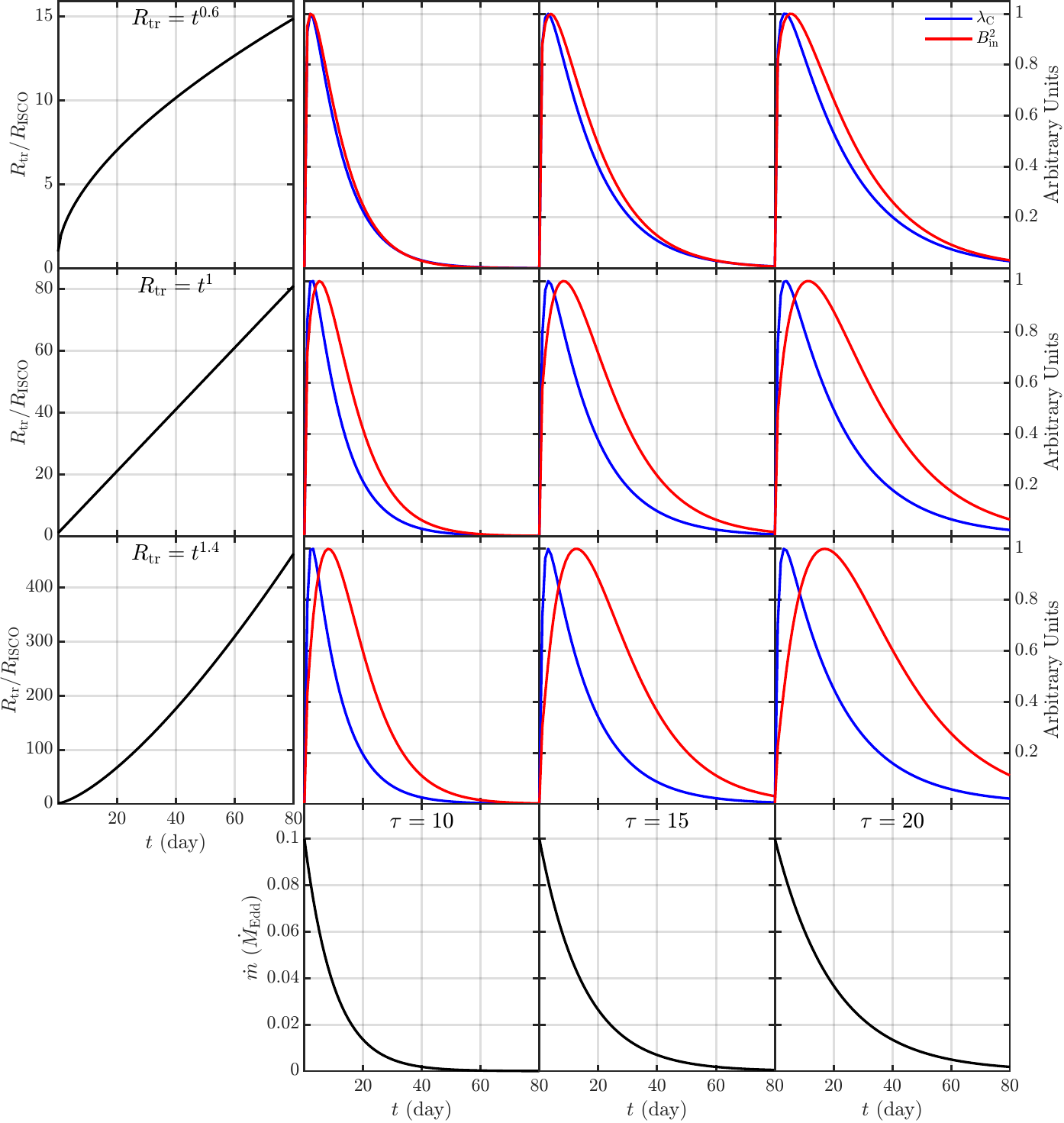}
	\caption{The temporal evolution of $\lambda_{\rm C}$ (blue; with $p_{\rm w}=0.05$) and $B_{\rm in}^2$ (red) is shown for nine combinations of the mass accretion rate $\dot{m}$ and the truncation radius $\Re$. The accretion rate decays exponentially with different timescales, i.e., $\dot{m} = 0.1 \exp(-t/\tau)$, while $\Re$ expands following a power-law. The corresponding evolutions of $\Re$ and $\dot{m}$ are presented in the first column and the bottom row, respectively, with each panel representing a specific combination. For instance, the panel in the second row and second column corresponds to the $\Re$ evolution shown in the second panel of the first column and the $\dot{m}$ evolution shown in the second panel of the bottom row. As we can see, a faster decay of $\dot{m}$ or a slower expansion of $\Re$ shortens the lag of $B_{\rm in}^2$ behind $\lambda_{\rm C}$. }
	\label{fig:e_d}
\end{figure*}

In the case of MAXI~J1348--630, no daily delay between the radio and X-ray emission is observed. This behavior may be attributed to a gradual expansion of the ADAF radius, such that the decay of $\dot{m}$ becomes the dominant factor in both $B_{\rm in}^2$ and $\lambda_{\rm C}$. To generally explain different observed behaviors, we consider three decay timescales for $\dot{m}$ and three power-law indices for $\Re$, yielding nine combinations plotted in Fig.~\ref{fig:e_d}. Here, $\dot{m}$ is assumed to decay exponentially, while $\Re$ grows according to a power law. For each combination, the temporal evolution of $\lambda_{\rm C}$ and $B_{\rm in}^2$ is shown. It is clear that either a more rapid decay of $\dot{m}$ or a slower expansion of $\Re$ reduces the delay of $B_{\rm in}^2$ behind $\lambda_{\rm C}$.

%Another supporting piece of evidence is that the radio flux decays more gradually than the Compton luminosity, which can be understood in the regime $\Re \gtrsim 20$, where $\lambda_{\rm C}$ depends only weakly on $\Re$, while $B_{\rm in}^2$ increases with $\Re$. Therefore, although the overall decay is primarily driven by the decrease in the mass accretion rate $\dot{m}$, $B_{\rm in}^2$ declines more gradually than $\lambda_{\rm C}$.

%In regim $\Re\gtrsim20$, $\lambda_{\rm C}$ is flat, however $B_{\rm in}^2$ increase with $\Re$, which can account for the observed gradual decay of the radio flux, compared with the Compton luminosity.

%In the rising hard state, which can be regarded as the inverse of the decay phase, the mass accretion rate, $\dot{m}$, which follows a broken evolution, rising to a peak before subsequently decaying, resulting various time-delay. 

%in the rising hard state, which can be regarded as the inverse of the decay phase. 
%the mass accretion rate $\dot{m}$ follows a broken temporal evolution: it increases rapidly to a peak and subsequently declines. This non-monotonic evolution naturally leads to various time-delay phenomena.
On the other hand, at the onset of the outburst, as $\dot{m}$ increases, the ADAF radius (i.e., the truncation radius) is large so that the increasing $\dot{m}$ dominates the evolution of both the Compton luminosity, $\lambda_{\rm C}$, and the inner magnetic field $B_{\rm in}$. This situation changes when the ADAF is sufficiently compact, e.g., when the truncation radius shrinks to $\Re \lesssim 20$.

\begin{figure*}
	\centering
    \includegraphics[width=1\linewidth]{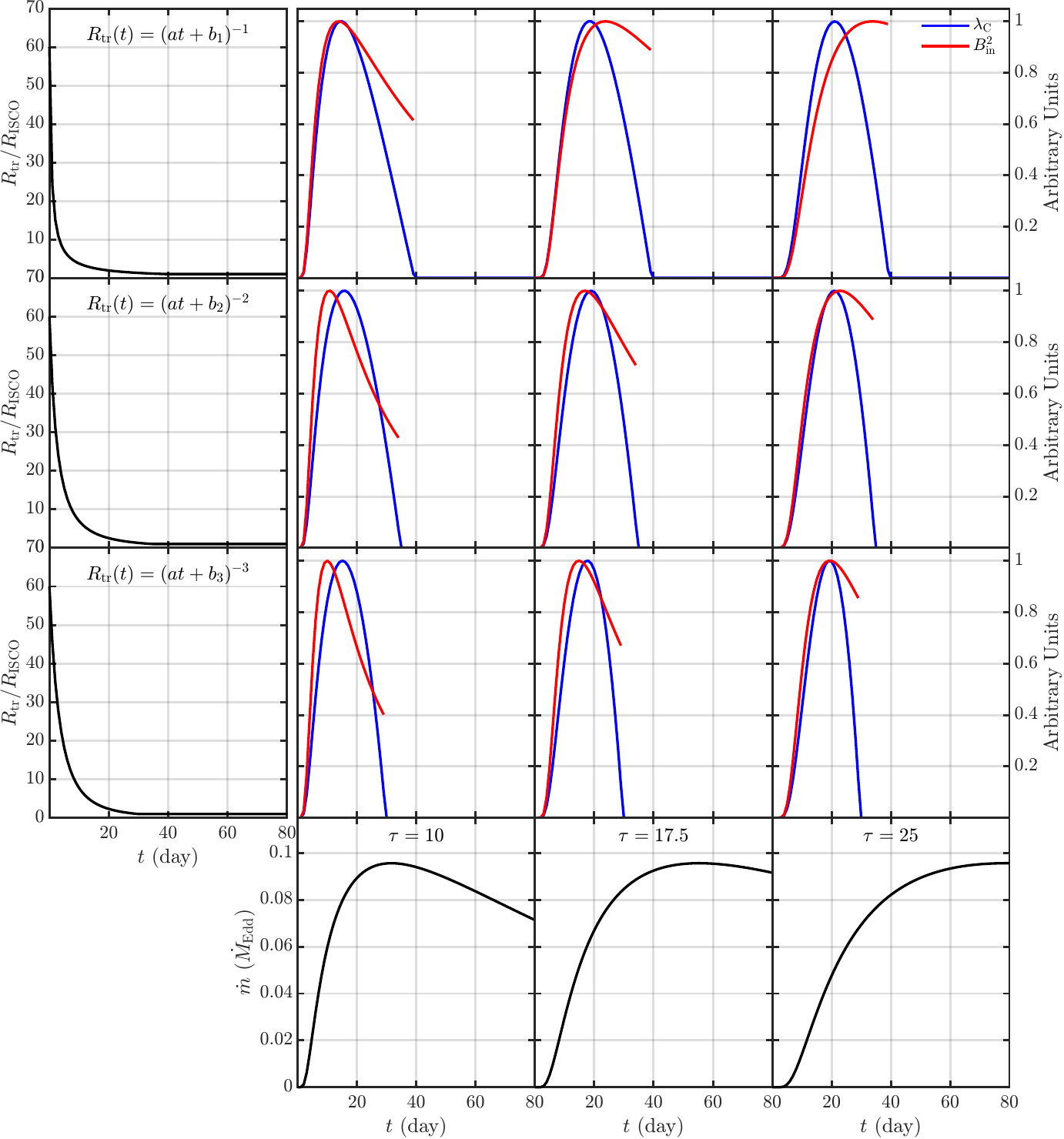}
	\caption{Temporal evolution of $\lambda_{\rm C}$ (blue; $p_{\rm w}=0.05$) and $B_{\rm in}^2$ (red) for nine combinations of the mass accretion rate $\dot{m}$ and truncation radius $\Re$. The corresponding evolutions of $\Re$ and $\dot{m}$ are shown in the first column and bottom row, respectively. 
    The accretion rate follows a fast-rise, exponential-decay (FRED) profile, given by $\dot{m}(t) = 0.2 \exp\!\left(-\frac{\tau}{t} - \frac{t}{10 \tau}\right)$, while $\Re$ evolves as a power law with different indices. We set $a=0.025$ and $\Re=60$ at $t=0$, yielding $b_1=0.017$, $b_2=0.13$, and $b_3=0.26$. 
    When $\Re<1$, we impose $\Re=1$, corresponding to the disappearance of the ADAF, in this case, $B_{\rm in}^2$ is terminated (producing a break), since the remaining thin disc cannot drive jets. This treatment does not affect the resulting time delay. 
  As shown, a later peak in $\dot{m}$ and a more rapid decrease of $\Re$, causing the $B_{\rm in}^2$ peak to occur after that of $\lambda_{\rm C}$. Conversely, in inverse cases, the $\lambda_{\rm C}$ peak can occur after the $B_{\rm in}^2$ peak. }
	\label{fig:e_r}
\end{figure*}

To explain the different behaviours observed in the rising hard state, we consider three rising curves for $\dot{m}$ and three power-law indices for $\Re$, yielding nine combinations shown in Fig.~\ref{fig:e_r}. The mass accretion rate $\dot{m}$ is assumed to follow a fast-rise, exponential-decay (FRED) profile, described by $\dot{m}(t) \propto \exp\!\left(-\frac{\tau}{t} - \frac{t}{10 \tau}\right)$, while the truncation radius $\Re$ decreases as a power law. For each combination, the temporal evolution of $\lambda_{\rm C}$ and $B_{\rm in}^2$ is shown.  

If $\dot{m}$ continues to increase rapidly when the truncation radius shrinks to $\Re \lesssim 20$, the Compton luminosity decreases sharply with decreasing $\Re$ in this regime, overwhelming the increase driven by $\dot{m}$. As a result, $\lambda_{\rm C}$ starts to decline earlier, whereas $B_{\rm in}^2$ decreases more gradually (see the first-row, last-column panel in Fig.~\ref{fig:e_r}). Consequently, the influence of $\Re$ on the evolution of $B_{\rm in}$ becomes dominant later than that for the Compton luminosity, producing a radio lag relative to the X-rays. This behavior is consistent with the observed radio lag during the rising hard state of MAXI~J1348--630, GRS~1739--278, and the second flare of H1743--322, where the corresponding peak of $\dot{m}$ may occur in the soft state.

However, an alternative scenario is possible in which the peak in $\dot{m}$ occurs during the hard state. In this case, at the beginning of the outburst, the mass accretion rate $\dot{m}$ increases exponentially and then gradually slows down, while the truncation radius shrinks in the regime $\Re \gtrsim 20$. In this case (an extreme scenario in which $\dot{m}$ remains nearly constant), the evolution of $B_{\rm in}$ is dominated by the decrease of $\Re$, causing $B_{\rm in}$ to decline first. In contrast, the Compton luminosity, $\lambda_{\rm C}$, remains slowly increasing with $\dot{m}$ because its dependence on $\Re$ is nearly flat in this regime. As a result, $B_{\rm in}$ reaches its peak earlier and subsequently begins to decline, while $\lambda_{\rm C}$ continues to rise and lags behind $B_{\rm in}$(see the third-row, second-column panel in Fig.~\ref{fig:e_r}), consistent with the observed Compton lag in GX~339--4 and H1743-322 first flare.

%In this phase, if $\dot{m}$ continues to increase exponentially while $r_{\rm tr}$ shrinks until it contracts to the ISCO, and the ADAF radius decreases only gradually when $r_{\rm tr} \lesssim 20$, then---because the Compton luminosity, $\lambda_{\rm C}$, decreases rapidly with decreasing $r_{\rm tr}$, $\lambda_{\rm C}$ reaches its peak earlier and subsequently begins to decline. Meanwhile, as $\dot{m}$ continues to increase exponentially, $B_{\rm in}^2$ keeps rising. As a consequence, $B_{\rm in}^2$ lags behind $\lambda_{\rm C}$, consistent with the observed radio lag.

%However, consider another extreme scenario in which $\dot{m}$ increases slowly and remains nearly constant, while $r_{\rm tr}$ gradually shrinks in the regime $r_{\rm tr} \gtrsim 20$. In this case, $B_{\rm in}^2$ decreases as $r_{\rm tr}$ shrinks, whereas the Compton luminosity, $\lambda_{\rm C}$, remains nearly constant. In a more general case, if $\dot{m}$ initially increases exponentially and then slows down or ceases before the truncation radius reaches $r_{\rm tr} \gtrsim 20$, $B_{\rm in}^2$ reaches its peak earlier and subsequently begins to decline, while $\lambda_{\rm C}$ continues to increase and lags behind $B_{\rm in}^2$, consistent with the observed Compton lag.

Overall, a later peak in $\dot{m}$ and a more rapid decrease of $\Re$, causing the $B_{\rm in}^2$ peak to occur after that of $\lambda_{\rm C}$. Conversely, in inverse cases, the $\lambda_{\rm C}$ peak can occur after the $B_{\rm in}^2$ peak. So, the diverse time-lag behaviors observed across different outbursts are governed by the coupled evolution of $R_{\rm tr}(t)$ and $\dot{m}(t)$, which may vary even within a single source. Elucidating the underlying accretion physics that dictates the functional relationship between $R_{\rm tr}$ and $\dot{m}(t)$ remains a subject for future research.

\section{Summary}
\label{sec:sum}
We conducted a time-delay analysis between the radio emission and the X-ray Compton luminosity during the 2010–2011 outburst of GX 339-4. Using the ICCF method, we examined the time delay between the Compton luminosity and the radio emission. Our main findings are as follows:

\begin{itemize}
    \item During the rising hard state, the Compton luminosity lags behind the radio emission by about 3 days.
    
    \item In contrast, during the decaying hard state, the radio emission lags behind the Compton luminosity by approximately 8 days, a delay also observed in MAXI J1820+070.
    
    \item During the decaying hard state, we estimate the inner magnetic field intensity to explain the radio delay by calculating the accretion rate $\dot{m}_r$ and the truncated radius $R_{\rm tr}$.
    
    \item During the rising hard state, by estimating the accretion rate $\dot{m}_r$ and truncated radius $R_{\rm tr}$, the calculated inner magnetic field intensity also explains the radio precede.

    \item By comparing the time delays observed in different outbursts across multiple sources, the physical origins of these differences are more clear.

\end{itemize}

%\end{appendix}

\section{Acknowledgements}
We thank S. Corbel for providing the radio data.
B.Y. is supported by Natural Science Foundation of China (NSFC) grants 12322307, 12361131579, 12273026, and 12373049; by “the Fundamental Research Funds for the Central Universities”; by Xiaomi Foundation / Xiaomi Young Talents Program. The data analysis in this paper have been done on the supercomputing system in the Supercomputing Center of Wuhan University.
X.W.C. is supported by NSFC grants 12533005 and 12233007; the science research grants from the China Manned Space Project with CMS-CSST-2025-A07

\clearpage
\clearpage
\bibliography{reference}{}
\bibliographystyle{aasjournal} % style aa.bst

\end{document}